\documentclass[preprintnumbers]{revtex4}
\UseRawInputEncoding
\usepackage{amssymb}
\usepackage{amsmath}
\usepackage{graphicx}
\usepackage{dcolumn}
\usepackage{bm}
\usepackage{subfigure}
\usepackage{color}

\begin{document}

\title{Restricted phase space thermodynamics of Einstein-power-Yang-Mills AdS black hole}
\author{Yun-Zhi Du$^1$, Huai-Fan Li$^1$\footnote{the corresponding author}, Yang Zhang$^1$, Xiang-Nan Zhou$^2$, and Jun-Xin Zhao$^3$}

\address{$^1$Institute of Theoretical Physics, Shanxi Datong University, Datong,  China\\
$^2$ College of Physics and Information Engineering, Shanxi Normal University, TaiYuan, China\\
$^3$ China Citic bank, China}

\thanks{\emph{e-mail:duyzh22@sxdtdx.edu.cn, huaifan999@163.com, zhangyphysics@126.com, zhouxn10@let.edu.cn, zhaojunxin163@163.com}}

\begin{abstract}
We consider the thermodynamics of the Einstein-Power-Yang-Mills AdS black holes in the context of the gauge-gravity duality. Under this framework, the Newton's gravitational constant and the cosmological constant are varied in the system. We rewrite the thermodynamical first law in a more extended form containing both the pressure and the central charge of the dual conformal field theory, i.e., the restricted phase transition formula. A novel phenomena arises: the dual quantity of pressure is the effective volume, not the geometric one. That is leading to a new behavior of the Van de Waals-like phase transition for this system with the fixed central charge: the supercritical phase transition. From the Ehrenfest's scheme perspective, we check out the second-order phase transition of the EPYM AdS black hole. Furthermore the effect of non-linear Yang-Mills parameter on these thermodynamical properties is also investigated.
\end{abstract}

%\pacs{04.70.Dy,05.70.Ce,04.20.Gz}
%\pacs{04.70.Dy 05.70.Ce}
\maketitle

\section{Introduction}
Since black holes and their thermodynamics can provide clues about the nature of quantum gravity, they have been of crucial importance. Especially, the asymptotically anti-de Sitter (AdS) black holes with a finite temperature can provide a description of the dual conformal field theory (CFT) via the AdS/CFT correspondence \cite{Maldacena1998}. Such black holes can be in thermal equilibrium with their radiation field and exhibit the Hawking-Page (HP) phase transition \cite{Hawking1983,Witten1998}. Subsequently, a very crucial point had been proposed: a negative cosmological constant can reduce a positive thermodynamical pressure, whose dual thermodynamical quantity is volume \cite{Kastor2009}. That makes AdS black holes can be identical to the ordinary thermodynamical systems, and their thermodynamics become more complete. In thus extended phase space, the mass parameter is interpreted as the entropy rather than the internal energy. And AdS black hole thermodynamics becomes more and more richer, such as the Van de Waals-like phase transition for the charged AdS black holes \cite{Kubiznak2012,Cai2013,Wei2015}, the reentrant phase transitions for the rotating system \cite{Altamirano2013,Frassin2014}, superfluid \cite{Hennigar2017a}, the polymer-like phase transition \cite{Dolan2014}, and the triple points \cite{Wei2014,Li2022}, along with the novel dual relation of HP phase transition \cite{Wei2020}. Meanwhile, the inclusion of the pressure-volume term in the thermodynamical first law makes other model parameters as novel thermodynamical quantities \cite{Cai2013} and make it possible to regard AdS black holes as heat engines \cite{Johnsom2014,Xu2017}. All of those developments are in the subdiscipline, black hole chemistry \cite{Kubiznak2017}.

On this issue, people always attempt to give the concrete physical explain of black holes chemistry in the extended phase space via AdS/CFT \cite{Dolan2014a,Kastor2014}. However it is somewhat elusive from the viewpoint of the holographic \cite{Zhang2015,Dolan2016}. For an AdS black hole system, the variations of the cosmological constant $\Lambda$ correspond to both the changing of the central charge and the CFT volume, which indicates that the thermodynamical first law in the extended phase space cannot be straightforwardly related to the corresponding thermodynamics of the dual field theory \cite{Karch2015,Sinamuli2017}. Furthermore the variation of pressure (or $\Lambda$) implies changing the gravity model. The corresponding ensemble does not describe the collection of black holes from the same gravity model where the macro states are the same, while describes the collection of gravity models of the same or similar black hole solutions. Additionally, in the extended phase space the absolute values of the coefficients appearing in the thermodynamical first law are not all one. These comments motivate the modification of the thermodynamical first law. Recently, authors in Refs. \cite{Cong2021,Visser2022} put forward the central charge and the chemical potential as a new pair of dual thermodynamical quantities which should be included in the thermodynamical first law. Here the Newton's gravitational constant can change as well as $\Lambda$, which will induce profound consequences of the chemical potential and its holographic interpretation. And compared with ordinary thermodynamical systems the introduction of these two thermodynamical quantities gives rise to a new thermodynamic phenomenon, the supercritical phase transition \cite{Gao2021,Zhao2022,Sadeghi2022}. In this work, for the Einstein-Power-Yang-Mills (EPYM) AdS black hole \cite{Zhang2015a,Corda2011,Mazharimousavi2009,Lorenci2002} we will exhibit the concrete process of establishing the more extended thermodynamical first law in which the pressure, volume, central charge, and chemical potential are included.

As we all know, at the linear level the charged black holes in an AdS spacetime nearby the critical point is of the scaling symmetries, $S\sim q^2,~P\sim q^{-2},~T\sim q^{-1}$ \cite{Johnson2018,Johnson2018a}, where $q$ is the electric charge. Whether the same scaling symmetry still hole on for the non-linear charged AdS black holes? There are lots of the generalization of the linear charged AdS black hole solution: Einstein-Maxwell-Yang-Mills AdS black hole \cite{Mazharimousavi2008}, Einstein-Power-Yang-Mills AdS black hole \cite{Lorenci2002}, Einstein-Maxwell-Power-Yang-Mills AdS black hole \cite{Zhang2015a}, Einstein-Yang-Mills-Gauss-Bonnet black hole \cite{Mazharimousavi2007}, Einstein-power-Maxwell-power-Yang-Mills-dilaton \cite{Stetsko2021}, and so on. An interesting non-linear generalization of charged black holes involves a Yang-Mill field exponentially coupled to Einstein gravity (i.e., Einstein-Power-Yang-Mills gravity theory) because it possesses the conformal invariance and is easy to construct the analogues of the four-dimensional Reissner-Nordstr\"{o}m black hole solutions in higher dimensions. Additionally several thermodynamical features of the EPYM AdS black hole in the extended phase space have been exhibited \cite{Zhang2015a,Du2021,Moumni2018}. Whether there appear new thermodynamical phenomena for the EPYM AdS black hole when the central charge and the chemical potential are introduced? The answer will be given in this work.

This work is organized as follows. In Sec. \ref{scheme2}, we briefly review the EPYM AdS black hole solution and its hawking temperature. In Sec. \ref{scheme3}, we derive the more extended thermodynamical first law that include the central charge and the chemical potential by considering the variation of the Newton's gravity constant and the cosmological constant.  Then, the critical thermodynamical quantities are exhibited and the effect of the non-linear YM parameter on the critical point is also investigated in Sec. \ref{scheme4}. In Sec. \ref{scheme5} we explore the the first-order phase transition in this more extended phase space are presented and compare the results with that in the extended phase space. Finally, we check out the Ehrenfest's scheme of the EPYM AdS black hole \ref{scheme6}. A brief summary is given in Sec. \ref{scheme7}.

\section{EPYM AdS black hole and Hawking temperature}
\label{scheme2}
The action for four-dimensional Einstein-power-Yang-Mills (EPYM) gravity with a cosmological constant $\Lambda$ was given by \cite{Zhang2015,Corda2011,Mazharimousavi2009,Lorenci2002}
\begin{eqnarray}
I=\frac{1}{2}\int d^4x\sqrt{g}
\left(R-2\Lambda-\mathcal{F}^\gamma\right)
\end{eqnarray}
with the Yang-Mills (YM) invariant $\mathcal{F}$ and the YM field $F_{\mu \nu}^{(a)}$
\begin{eqnarray}
\mathcal{F}&=&\operatorname{Tr}(F^{(a)}_{{\mu\nu}}F^{{(a)\mu\nu}}),\\
F_{\mu \nu}^{(a)}&=&\partial_{\mu} A_{\nu}^{(a)}-\partial_{\nu} A_{\mu}^{(a)}+\frac{1}{2 \xi} C_{(b)(c)}^{(a)} A_{\mu}^{(b)} A_{\nu}^{(c)}.
\end{eqnarray}
Here, $\operatorname{Tr}(F^{(a)}_{\mu\nu}F^{(a)\mu\nu})
=\sum^3_{a=1}F^{(a)}_{\mu\nu}F^{(a)\mu\nu}$, $R$ and $\gamma$ are the scalar curvature and a positive real parameter, respectively; $C_{(b)(c)}^{(a)}$ represents the structure constants of three-parameter Lie group $G$; $\xi$ is the coupling constant; and $A_{\mu}^{(a)}$ represents the $SO(3)$ gauge group Yang-Mills (YM) potentials defining by the Wu-Yang (WY) ansatz \cite{Balakin2016}. Variation of the action with respect to the spacetime metric $g_{\mu\nu}$ yields the field equations
\begin{eqnarray}
&&G^{\mu}{ }_{\nu}+\Lambda\delta^{\mu}{ }_{\nu}=T^{\mu}{ }_{\nu},\\
&&T^{\mu}{ }_{\nu}=-\frac{1}{2}\left(\delta^{\mu}{ }_{\nu} \mathcal{F}^{\gamma}-4 \gamma \operatorname{Tr}\left(F_{\nu \lambda}^{(a)} F^{(a) \mu \lambda}\right) \mathcal{F}^{\gamma-1}\right).
\end{eqnarray}
Variation with respect to the 1-form YM gauge potentials $A_{\mu}^{(a)}$ and implement the traceless yields the 2-forms YM equations
\begin{equation}
\mathbf{d}\left({ }^{\star} \mathbf{F}^{(a)} \mathcal{F}^{\gamma-1}\right)+\frac{1}{\xi} C_{(b)(c)}^{(a)} \mathcal{F}^{\gamma-1} \mathbf{A}^{(b)} \wedge^{\star} \mathbf{F}^{(c)}=0,
\end{equation}
where $\mathbf{F}^{(a)}=\frac{1}{2}F_{\mu \nu}^{(a)}dx^\mu\wedge dx^\nu,~\mathbf{A}^{(b)}=A_{\mu }^{(b)}\wedge dx^\mu$, and ${ }^{\star}$ stands for the duality. It is obviously that for the case of $\gamma=1$ the EPYM theory reduces to the standard Einstein-Yang-Mills (EYM) theory \cite{Mazharimousavi2007a}. In this work our issue is paid on the role of the non-linear YM charge parameter $\gamma$.

Here we should point out that the non-Abelian property of the YM gauge field is expressed with its YM potentials
\begin{eqnarray}
\mathbf{A}^{(b)}=\frac{q}{r^2}C^{(a)}_{(i)(j)}x^idx^j,~r^2=\sum_{j=1}^3x_j^2,
\end{eqnarray}
and $q$ is the YM charge, the indices ($a,~i,~j$) run the following ranges: $1\leq a,~i,~j\leq3$. The coordinates $x_i$ take the following forms: $x_1=r \cos\phi \sin\theta,~x_2=r \sin\phi \sin\theta,~x_3=r \cos\theta.$ Since we have utilized the WY ansatz for the YM field, the invariant for this field takes the form \cite{Stetsko2020,Chakhchi2022}
\begin{eqnarray}
\operatorname{Tr}(F^{(a)}_{{\mu\nu}}F^{{(a)\mu\nu}})=\frac{q^2}{r^4}.
\end{eqnarray}
This form leads to the disappearance of the structure constants which can be described the non-Abelian property of the YM gauge field. Therefore, under the condition of the WY ansatz we may focus on the role of the non-linear YM charge parameter, instead of the non-Abelian character parameter.

The metric for the four-dimensional EPYM AdS black hole is given as follows \cite{Yerra2018},
\begin{eqnarray}
d s^{2}=-f(r) d t^{2}+f^{-1} d r^{2}+r^{2} d \Omega_{2}^{2},
\end{eqnarray}
where
\begin{eqnarray}
%f(r)=\begin{cases}
%1-\frac{2GM}{r}+\frac{r^{2}}{l^2}+\frac{G\left(2q^{2}\right)^{\gamma}}{2(4 \gamma-3) r^{4 \gamma-2}}, & ~~~~~~~~~~\gamma\neq\frac{3}{4}\\
%1-\frac{2GM}{r}+\frac{r^{2}}{l^2}-\frac{G q^{3/2}\ln r}{2^{1/4}r}. & ~~~~~~~~~~\gamma=\frac{3}{4}
%\end{cases}
f(r)=1-\frac{2GM}{r}+\frac{r^{2}}{l^2}+\frac{G\left(2q^{2}\right)^{\gamma}}{2(4 \gamma-3) r^{4 \gamma-2}}. \label{f}
\end{eqnarray}
Here $d\Omega_{2}^{2}$ is the metric on unit $2$-sphere with volume $4\pi$ and $q$ is the YM charge, $l$ is related to the cosmological constant: $l^2=-\frac{3}{\Lambda}$, $\gamma$ is the non-linear YM charge parameter and satisfies $\gamma>0$ \cite{Corda2011}. The event horizon of the black hole is obtained from the relation $f(r_+)=0$. The mass parameter of the black hole can be expressed in terms of the horizon radius as
\begin{eqnarray}
M=\frac{r_+}{2G}\left(1+\frac{r^2_+}{l^2}+\frac{2^{\gamma-1}Gq^{2\gamma}}{(4\gamma-3)r_+^{4\gamma-2}}\right).\label{M}
\end{eqnarray}
We can also obtain the Hawking temperature of the black hole from eq. (\ref{f}) as follows
\begin{eqnarray}
T=\frac{1}{4 \pi r_{+}}\left(1+8 \pi GP r_{+}^{2}-\frac{G\left(2 q^{2}\right)^{\gamma}}{2 r_{+}^{(4 \gamma-2)}}\right).\label{T}
\end{eqnarray}
From eqs. (\ref{M}) and (\ref{T}) we will calculate the critical value of thermodynamical quantities which are presented in Sec. VI. Next we will give the modified first law of the four-dimensional EPYM AdS black hole thermodynamics in natural units ($\hbar=c=1$), i.e., the restricted phase space formulism.

\section{Restricted phase space formulism of EPYM AdS black hole}
\label{scheme3}

Recently people in Refs. \cite{Kastor2009,Kubiznak2012} proposed that the negative cosmological constant could induce a positive thermodynamical pressure, which is in terms of the cosmological constant and the Newton's gravitational constant as
\begin{eqnarray}
P=-\frac{\Lambda}{8\pi G}~~~~\text{or}~~~~P=\frac{3}{8\pi G l^2}.  \label{PP}
\end{eqnarray}
In the above equation, the pressure will change with the variation of the cosmological constant and the Newton's gravitational constant. In natural units, the Bekenstein-Hawking entropy reads
\begin{eqnarray}
S=\frac{A}{4G}=\frac{\pi r_+^2}{G}, \label{SS}
\end{eqnarray}
where $A$ is the area of the black hole. The Hawking temperature can be expressed in the term of the surface gravity $\kappa$ as
\begin{eqnarray}
T=\frac{\kappa}{2\pi}. \label{TT}
\end{eqnarray}
In the extended phase space, the Newton's gravitational constant is fixed and the mass of black hole is interpreted as the enthalpy instead of the internal energy. Thus the general form of the thermodynamical first law for the EPYM AdS black hole of the surface gravity, charge, the cosmological constant, and the area are
\begin{eqnarray}
\delta M&=&T\delta S+V\delta P+\Psi\delta q^{2\gamma}\nonumber\\
&=&\frac{\kappa}{8\pi G}\delta A-\frac{V}{8\pi G}\delta\Lambda+\Psi\delta q^{2\gamma},\label{deltaM}
\end{eqnarray}
where the volume and potential are
\begin{eqnarray}
V=\frac{4\pi r_+^3}{3},~~~~
\Psi=
\frac{2^{\gamma-2}}{(4\gamma-3)r_+^{4\gamma-3}}.\label{VPsi}
%\Psi=\begin{cases}
%\frac{2^{\gamma-2}}{(4\gamma-3)r_+^{4\gamma-3}}& ~~~~~~~~~~\gamma\neq\frac{3}{4}\\
%-2^{-5/4}\ln r_+& ~~~~~~~~~~\gamma=\frac{3}{4}
%\end{cases}.
\end{eqnarray}
We can check out the final expression in eq. (\ref{deltaM}) by using eqs. (\ref{PP}), (\ref{SS}), and (\ref{TT}). In the expanded phase space, the thermodynamical phase transition properties of the EPYM AdS black hole were exhibited in Refs. \cite{Du2021} and the corresponding optical properties including the photon sphere and shadow were also presented in Refs. \cite{Du2022,Du2022a}. Next we will exhibited the concrete details of the restricted phase space formulism for the EPYM AdS black hole in the case of $\gamma\neq 3/4$.

It has been shown that the Holographic interpretation of the above thermodynamical first law in eq. (\ref{deltaM}) could cause some issues \cite{Johnsom2014,ZhangJHEP2015,McCarthyJHEP2017}. The $V\delta P$ (the variation of the cosmological constant) in the thermodynamical first law of the bulk is shown to have two terms in the thermodynamical first law at the boundary conformal field theory (CFT): one is the central charge of the boundary CFT and the thermodynamical pressure of the boundary CFT which is caused by the change of the AdS radius. The way of addressing this problem is to invoke the form of the central charge from the AdS/CFT dictionary, which is related to the AdS radius $l$ as in Ref. \cite{Karch2015}
\begin{eqnarray}
C=\frac{kl^2}{16\pi G}.\label{C}
\end{eqnarray}
Here the parameter $k$ is determined on the details of the system on the boundary. In order to keep $C$ as a constant, we should vary the Newton's gravitational constant $G$ as well as the AdS radius $l$. When considering the mass parameter $M$ to be a function of the area $A$, the cosmological constant $\Lambda$, the charge $q^{2\gamma}$, and the Newton's gravitational constant $G$, i.e., $M\equiv M(A,~\Lambda,~q^{2\gamma},~G)$, the variation of $M$ can be rewritten as
\begin{eqnarray}
\delta M=\frac{\partial M}{\partial A}\delta A+\frac{\partial M}{\partial \Lambda}\delta \Lambda+\frac{\partial M}{\partial q^{2\gamma}}\delta q^{2\gamma}+\frac{\partial M}{\partial G}\delta G.  \label{deltaMM}
\end{eqnarray}
Compared with eq. (\ref{deltaM}) we can see that the conjugate variables of $A$, $\Lambda$, $q^{2\gamma}$ are $\frac{\kappa}{8\pi G},~-\frac{V}{8\pi G},~\Psi$. With the definition $G\frac{\partial M}{\partial G}=-\xi$ we can recast the above equation as
\begin{eqnarray}
\delta M=\frac{\kappa}{8\pi G}\delta A-\frac{V}{8\pi G}\delta\Lambda+\Psi\delta q^{2\gamma}-\xi\frac{\delta G}{G}.\label{deltaMG}
\end{eqnarray}
In the following, we try to give the coefficient $\xi$ in the above equation. For that we make use of a modified mass term as suggested in \cite{Cong2021}
\begin{eqnarray}
GM=\mathcal{M}(A,\Lambda,Gq^{2\gamma}).\label{GM}
\end{eqnarray}
Performing the differential of the above equation and combining eq. (\ref{deltaMM}), we have
\begin{eqnarray}
G\delta M&=&\frac{\partial \mathcal{M}}{\partial A}\delta A+\frac{\partial \mathcal{M}}{\partial \Lambda}\delta \Lambda+\frac{G\partial \mathcal{M}}{\partial (Gq^{2\gamma})}\delta q^{2\gamma}+\left(\frac{q^{2\gamma}\partial \mathcal{M}}{\partial (Gq^{2\gamma})}-M\right)\delta G,\\
\Rightarrow ~~~~\delta M&=&\frac{\partial \mathcal{M}}{G\partial A}\delta A+\frac{\partial \mathcal{M}}{G\partial \Lambda}\delta \Lambda+\frac{\partial \mathcal{M}}{\partial (Gq^{2\gamma})}\delta q^{2\gamma}+\left(\frac{q^{2\gamma}\partial \mathcal{M}}{\partial (Gq^{2\gamma})}-M\right)\frac{\delta G}{G}. \label{deltaM0}
\end{eqnarray}
Comparing the above equation with eq. (\ref{deltaMG}), we can obtain the following expressions
\begin{eqnarray}
\frac{\partial \mathcal{M}}{\partial A}=\frac{\kappa}{8\pi},~~~~\frac{\partial \mathcal{M}}{\partial \Lambda}=-\frac{V}{8\pi},~~~~\frac{\partial \mathcal{M}}{\partial (Gq^{2\gamma})}=\Psi,~~~~\frac{q^{2\gamma}\partial \mathcal{M}}{\partial (Gq^{2\gamma})}-M=-\xi.
\end{eqnarray}
Therefore the coefficient $\xi$ has the form as
\begin{eqnarray}
\xi=M-q^{2\gamma}\Psi.
\end{eqnarray}
Performing the differential of the cosmological constant, the pressure in eq. (\ref{PP}), the area in eq. (\ref{SS}), and the central charge in eq. (\ref{C}), we have
\begin{eqnarray}
\frac{\delta\Lambda}{\Lambda}=-\frac{\delta C}{2C}+\frac{\delta P}{2P},~~
\frac{\delta G}{G}=-\left(\frac{\delta C}{2C}+\frac{\delta P}{2P}\right),~~
\frac{\delta A}{A}=\frac{\delta S}{S}-\frac{\delta P}{2P}-\frac{\delta C}{2C},
\end{eqnarray}
Combining these results, the differential of the black hole mass in eq. (\ref{deltaM0}) can be rewritten as a function of the thermodynamical quantities ($~S,~P,~q^{2\gamma},~C$)
\begin{eqnarray}
\delta M=T\delta S+V_{eff}\delta P+\Psi\delta q^{2\gamma}+\mu\delta C,
\end{eqnarray}
where $V_{eff}$ and $\mu$ are the effective thermodynamical volume and the chemical potential and they have the following forms
\begin{eqnarray}
V_{eff}&=&\frac{1}{2P}\left(M-TS+PV-q^{2\gamma}\Psi\right),\\
\mu&=&\frac{1}{2C}\left(M-TS-PV-q^{2\gamma}\Psi\right).
\end{eqnarray}
That is the restricted phase space formulism of the four-dimensional EPYM AdS black hole in the case of $\gamma\neq3/4$.
For convenience, we introduce the induced thermodynamical quantities as
\begin{eqnarray}
\bar M=GM,~~\bar S=GS,~~\bar P=GP,~~\bar q=Gq^{2\gamma},~~\bar C=GC,~~
\end{eqnarray}
the thermodynamical first law in the restricted phase space becomes
\begin{eqnarray}
\delta\bar M=T\delta\bar S+V_{eff}\delta\bar P+\Psi\delta\bar q+\mu\delta \bar C, \label{deltabarM}
\end{eqnarray}
and the effective volume and the chemical potential are
\begin{eqnarray}
V_{eff}&=&\frac{1}{2\bar P}\left(\bar M-T\bar S+\bar PV-\frac{\bar q\Psi}{2}\right),\\
\mu&=&\frac{1}{2\bar C}\left(\bar M-T\bar S-\bar P V-\frac{\bar q\Psi}{2}\right).
\end{eqnarray}
From above equations and eq. (\ref{VPsi}), we can see that the Euler relation of EPYM AdS black holes in the restricted phase space is indeed restored as in an ordinary thermodynamical system, which reads
\begin{eqnarray}
\bar M=T\bar S+\bar P V_{eff}+\bar q\Psi+\mu\bar C.
\end{eqnarray}
In the following we will use these induced quantities ($\bar S,~\bar P,~\bar q,~\bar C$) and $T,~V_{eff},~\Psi,~\mu$ to investigate the thermodynamical properties of this system.

\section{Critical curves of EPYM AdS black hole}
\label{scheme4}

\begin{figure}[htp]
\centering
\subfigure[$T_c-\bar P_c$]{\includegraphics[width=0.4\textwidth]{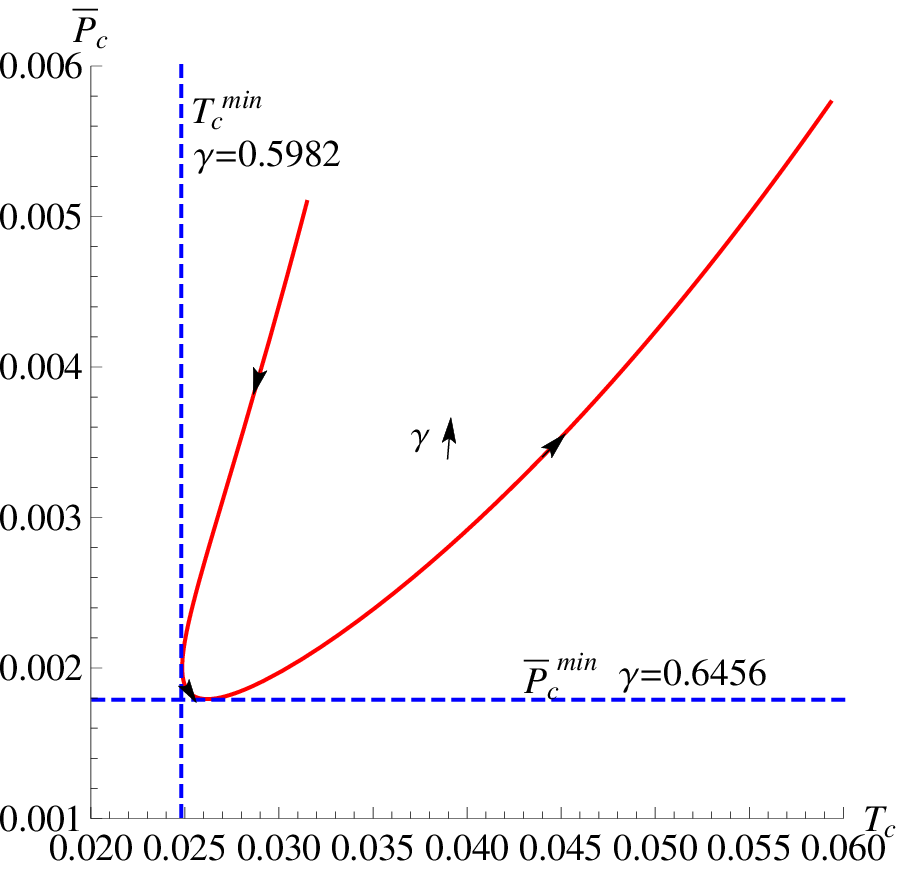}\label{Tc-Pc-gamma}}~~~~
%\subfigure[$\mu_c-C_c$]{\includegraphics[width=0.4\textwidth]{muc-Cc-gamma.eps}\label{Tc-Pc-gamma}}~~~~\\
\subfigure[$T_c-\bar C_c$]{\includegraphics[width=0.4\textwidth]{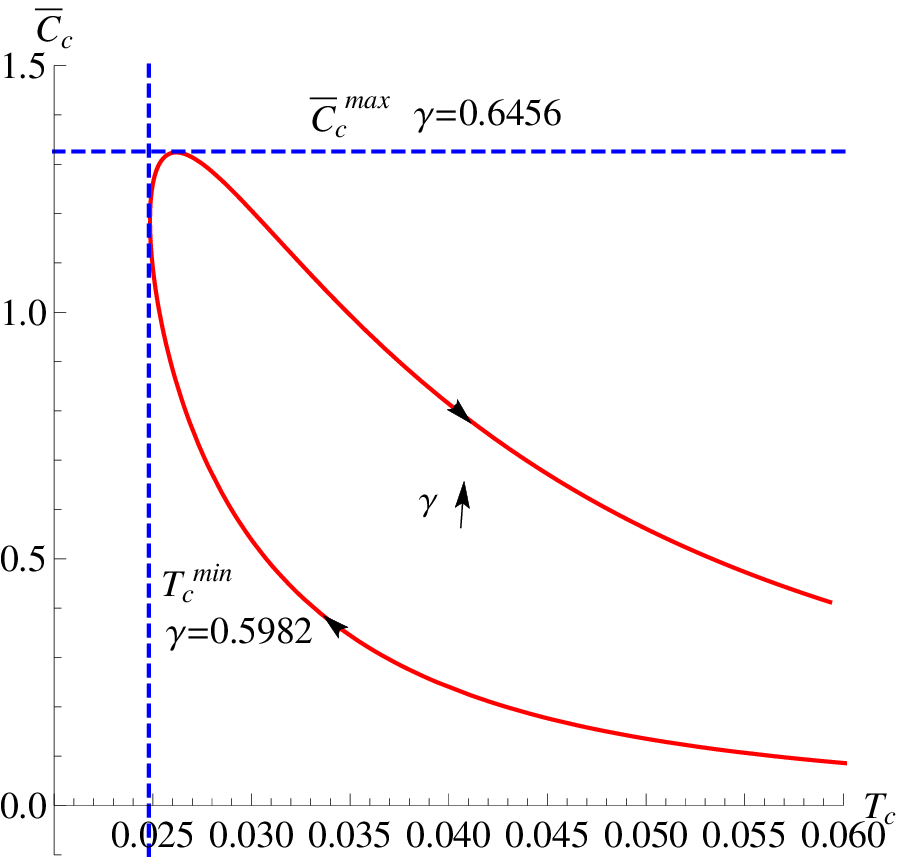}\label{Tc-Cc-gamma}}
\caption{The critical curves in $T_c-\bar P_c$ and $T_c-\bar C_c$ diagrams with the non-linear charge parameter $\gamma$. The YM charge is set to $\bar q=1$.}\label{Tc-Pc-Cc-gamma}
\end{figure}

Based on the classification of phase transition for a thermodynamical system by Ehrenfest, the critical point can be obtained by the following equations
\begin{eqnarray}
\frac{\partial T}{\partial \bar S}=\frac{\partial^2T}{\partial \bar S^2}=0.
\end{eqnarray}
With eqs. (\ref{T}) and (\ref{SS}) and above equations, we have
\begin{eqnarray}
r_{c}^{4\gamma-2}=\gamma(4\gamma-1)2^{\gamma}\bar q,~~~~l_{c}^2=\frac{6\gamma}{2\gamma-1}r_{c}^2.
\end{eqnarray}
The other critical quantities in the restricted phase space are
\begin{eqnarray}
\bar P_c&=&\frac{2 \gamma -1}{16 \pi\gamma  r_{c}^2},~~V^{c}_{eff}=\frac{1}{3} \pi  r_c^3 \left(\frac{6 \gamma  }{2 \gamma -1}+\frac{3\gamma2^{\gamma+1} \bar q r_c^{2-4 \gamma }}{4 \gamma -3}+1\right),\\
T_c&=&\frac{2\gamma -1}{(4\gamma -1)\pi  r_{c}},~~
\bar S_c=\pi r_{c}^2,\\
\bar C_c&=&\frac{3 \gamma  r_c^2}{8 \pi  (2 \gamma -1)},~~
\mu_c=\frac{\pi  (2 \gamma -1) }{6 \gamma ^2 r_c}\left(\frac{2^{\gamma +1} \gamma  (2 \gamma -1) \bar q r_c^{2-4 \gamma }}{4 \gamma -3}+1\right)
\end{eqnarray}
The results indicate that the critical point is determined by the non-linear YM parameter $\gamma$ and the YM charge $\bar q$. Since the effect of $\bar q$ on the phase transition have been investigated in the previous works \cite{Du2021,Du2022}, here we only exhibit the effects of $\gamma$ on the critical temperature, the critical pressure, and the critical central charge in Fig. \ref{Tc-Pc-Cc-gamma}. In the range $0.5<\gamma\le0.5982$, the critical temperature and pressure are both decreasing with the increasing of $\gamma$, while the critical central charge is increasing. When $0.6456\le\gamma$, the critical temperature and pressure are the monotonically increasing functions with $\gamma$, while the critical central charge is not. In the middle range $0.5982\le\gamma\le0.6456$, the critical central charge and pressure is increasing, while the critical temperature is decreasing.
%%%%%%%%%%%%%%%%%%%%%%%%%%%%%%%%%%%%%%%%%%%%%%%%%%%%%%%%%%%%%%%%%%%%%%%%%%%%%%%%%%%%%%%%%%%%%%%%%%

\section{First-order phase transition in restricted phase space}
\label{scheme5}
In the previous works \cite{Du2021,Du2022}, the phase transition condition of the EPYM AdS black hole in the extended phase space had been proposed. Furthermore the first-order phase diagrams of $T-S$, $P-V$, and $q^{2\gamma}-\Psi$ in the extended phase space also were exhibited. In this manuscript we introduce the central charge and the chemical potential to present the phase structure of the EPYM AdS black hole in the restricted phase space, where the volume is modified. Thus we mainly focus on the phase diagrams in the $T-\bar S$, $\bar P-V_{eff}$, and $\bar C-\mu$ planes, and exhibit the corresponding properties of phase transition. Firstly, we will review the phase transition condition from the viewpoint of the independent dual thermodynamical quantities $T-\bar S$.

For the EPYM black hole with the given YM charge $\bar q$ and pressure $\bar P_0<\bar P_c$, in the phase diagram of $T-\bar S$ the entropies at the boundary of the two-phase coexistence area are remarked by $\bar S_1$ and $\bar S_2$, respectively. The corresponding phase transition temperature is $T_0$, which is related with the horizon radius $r_+$. Therefore, from the Maxwell's equal-area law $T_0(\bar S_2-\bar S_1)=\int^{\bar S_2}_{\bar S_1}Td\bar S$ and eq. (\ref{T}), we have
\begin{eqnarray}
2 \pi T_{0}=\frac{1}{r_{2}(1+x)}+\frac{8 \pi \bar P_0 r_{2}}{3(1+x)}\left(1+x+x^{2}\right)-\frac{2 ^{\gamma}\bar q r_{2}^{1-4 \gamma}}{2(3-4 \gamma)} \frac{\left(1-x^{3-4 \gamma}\right)}{\left(1-x^{2}\right)}\label{TEAL}
\end{eqnarray}
with $x=\frac{r_1}{r_2}$. In addition, from the state equation we have
\begin{eqnarray}
T_{0}=\frac{1}{4 \pi r_{1,2}}\left(1+8 \pi \bar P_0 r_{1,2}^{2}-\frac{2^{\gamma} \bar q}{2 r_{1,2}^{(4 \gamma-2)}}\right),~~~~\label{T01}
\end{eqnarray}
From two above equations, we have
\begin{eqnarray}
0&=&-\frac{1-x}{r_{2} x}+8 \pi \bar P_0 r_{2}(1-x)+\frac{2^{\gamma} \bar q}{2 r_{2}^{4 \gamma-1} x^{4 \gamma-1}}\left(1-x^{4 \gamma-1}\right),\label{T000}~~~~\\
8 \pi T_{0}&=&\frac{1+x}{r_{2} x}+8 \pi \bar P_0 r_{2}(1+x)-\frac{2^{\gamma} \bar q}{2 r_{2}^{4 \gamma-1} x^{4 \gamma-1}}\left(1+x^{4 \gamma-1}\right).\label{T001}
\end{eqnarray}
Considering eqs. (\ref{TEAL}), (\ref{T000}), and (\ref{T001}), the horizon $r_2$ has the following form
\begin{eqnarray}
r_{2}^{4 \gamma-2}=\frac{2^{\gamma} \bar q\left[(3-4 \gamma)(1+x)\left(1-x^{4 \gamma}\right)+8 \gamma x^{2}\left(1-x^{4 \gamma-3}\right)\right]}{2 x^{4 \gamma-2}(3-4 \gamma)(1-x)^{3}}
=2^{\gamma} \bar q f(x,\gamma).
\label{r2}
\end{eqnarray}
\begin{figure}[htp]
\centering
\subfigure[$\gamma=0.85,~\bar q=1$]{\includegraphics[width=0.4\textwidth]{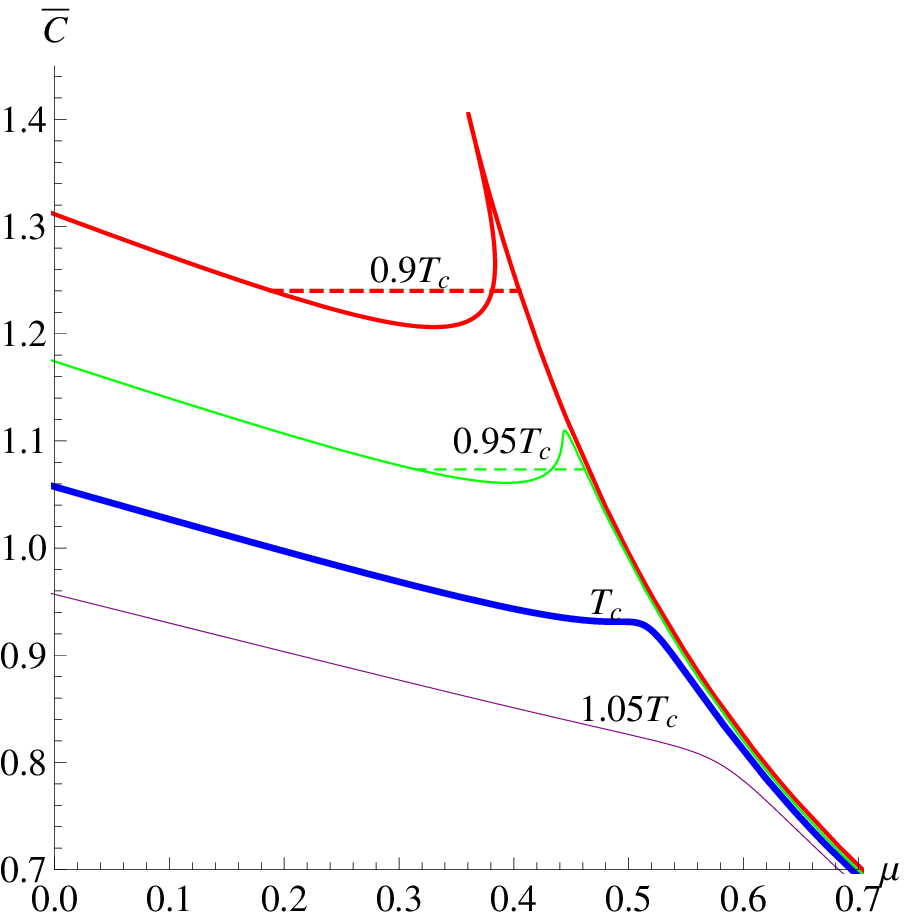}\label{muC185}}~~~~
\subfigure[$\gamma=0.85,~\bar q=1$]{\includegraphics[width=0.4\textwidth]{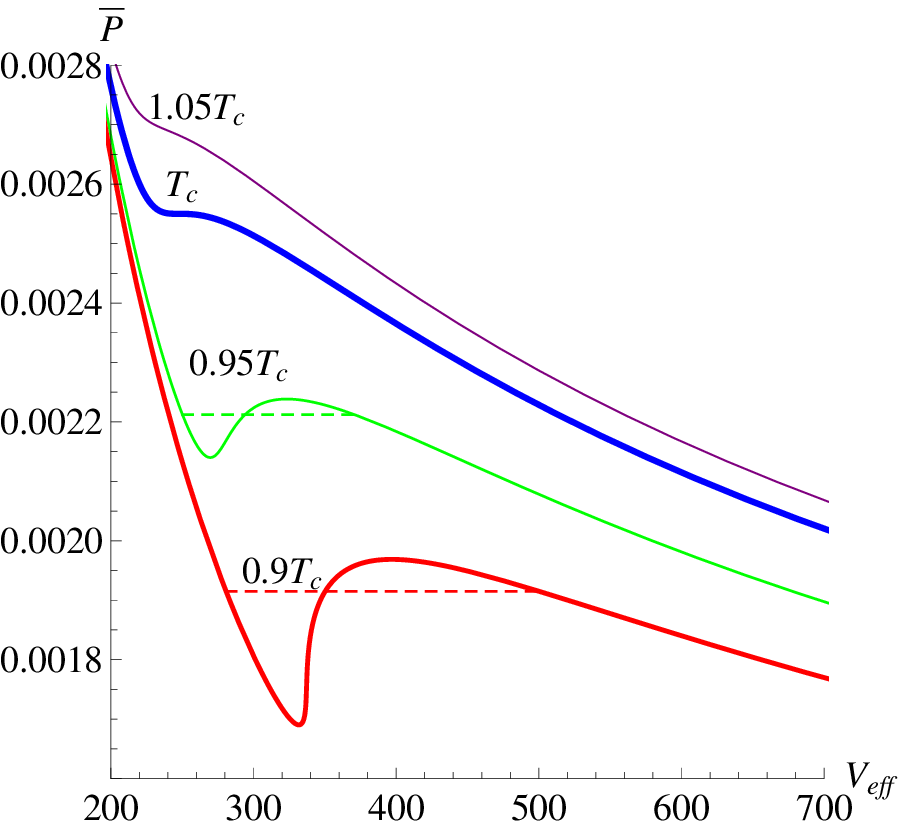}\label{Veff-P185}}
\caption{The phase diagrams of $V_{eff}-\bar P$ and $\mu-\bar C$ with different values of temperature.}\label{nu-c-v-p}
\end{figure}
\begin{figure}[htp]
\subfigure[$\gamma=1,~\bar q=1$]{\includegraphics[width=0.4\textwidth]{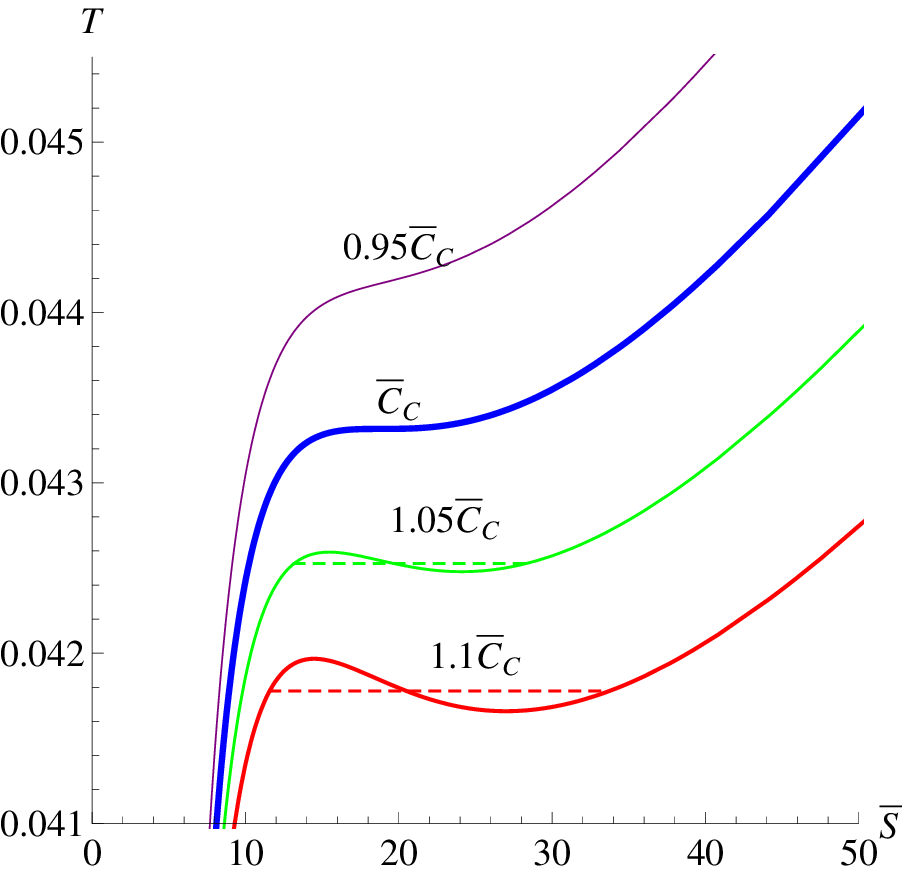}\label{T-S-C}}~~~~
\subfigure[$\gamma=0.85,~\bar q=1$]{\includegraphics[width=0.4\textwidth]{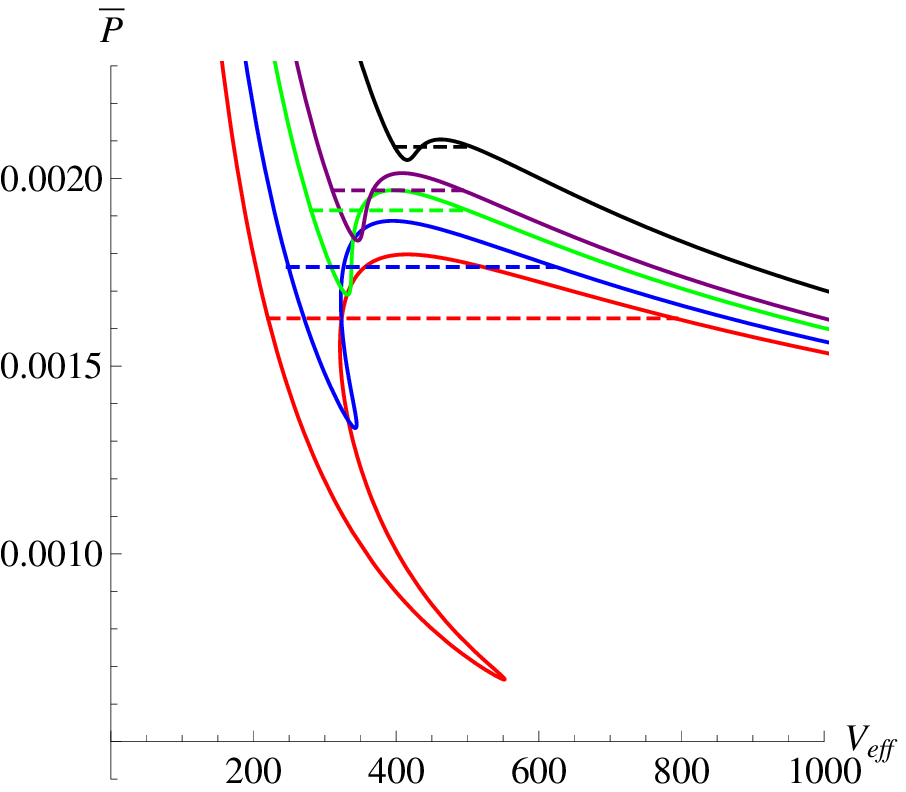}\label{Veff-P-gamma}}
\caption{The first-order phase diagrams of $V_{eff}-\bar P$ with different values of temperature. The parameters are set to $\bar q=1,~T=0.033<T_c$. And the non-linear YM charge parameter varies from 0.8 to 1 from the black line to the red one.}\label{T-S-C-V-P-gamma}
\end{figure}
Since the horizon radius must be positive, the non-linear YM charge parameter satisfies the condition $\frac{1}{2}<\gamma$ and $\gamma\neq\frac{3}{4}$. In addition, from the state equation (\ref{T01}) the temperature $T_0$ can be written as a function of $r_+$ and $x$
\begin{eqnarray}
T_0=\frac{1}{4\pi x(1-x)r_2}\left(1-x^2-\frac{2^{\gamma-1}\bar q(1-x^{4\gamma})}{(x r_2)^{4\gamma-2}}\right).\label{T0000}
\end{eqnarray}
Considering eqs. (\ref{r2}) and (\ref{T0000}), we find that for the given values of $\gamma$, $\bar q$, and temperature $T_0$, we can calculate the value of $x$. Then substituting the result of $x$ into eq. (\ref{r2}) the large horizon radius will be obtained. Thus for the given temperature $T_0$ ($T_0<T_c$), the first-order phase transition condition reads
\begin{eqnarray}
\frac{\left(2 q^{2}\right)^{\gamma}}{r_{2}^{4 \gamma-2}}=\frac{1}{f(x, \gamma)}. \label{PTC}
\end{eqnarray}
In other words, the phase transition of this system is determined by the radio between the YM charge $\bar q$ and $r_{2}^{4 \gamma-2}$, not just only the black hole horizon. Note that we call this radio as the YM potential at the horizon surface $r_2$. Thus we call this phase transition as the high/low-potential black hole (HPBH/LPBL) one.

The phase diagrams of $\bar P-V_{eff}$, $\bar C-\mu$, and $T-\bar S$ are shown in Figs. \ref{nu-c-v-p} and \ref{T-S-C}. It is very interesting that for the system with the higher central charge ($\bar C>\bar C_c$) the first-order phase transition appears, while it will vanish as $\bar C<\bar C_c$. This phenomena is consistent with that in Ref. \cite{Cong2021} which is called as the supercritical phase transition and governed by the freedom degree in the conformal field theory. However it is fully different from the system undergoing the isobaric processes in the extended phase space \cite{Du2021} as well as the another kind of supercritical phase transition \cite{Gao2021} where the central charge and chemistry potential exist while not the pressure and volume. Therefore, it can be guessed that the effects of pressure and central charge on the first-order phase transition are completely opposite. In addition, although the volume in the restrict phase space is modified compared with that in the expanded phase space, the phase transition point is still the same in the volume-pressure plane. And for the lower temperature (less than the critical one), there exist the first-order phase transition both in the $\bar P-V_{eff}$ and $\bar C-\mu$ planes, for the higher temperature the first-order phase transition of the system disappear. However, there exists a very interesting phenomena:  when the central charge of this system is low than the critical one, the phase transition does not exist, while for the higher central charge it is appearing. That is fully different from other thermodynamical quantities. In addition the effect of the non-linear YM parameter on the phase diagram of $\bar P-V_{eff}$ has been shown in Fig.  \ref{Veff-P-gamma}.

\section{Phase transition From Ehrenfest's equations}
\label{scheme6}
We now exploit Ehrenfest's scheme in order to understand the nature of the phase transition. Ehrenfest's scheme basically consists of a pair of equations known as Ehrenfest's equations of first and second kind. For a standard thermodynamic system these equations may be written as \cite{Banerjee004,Banerjee156,Banerjee317}
\begin{eqnarray}
\left(\frac{\partial\bar P}{\partial T}\right)_{\bar S}&=&\frac{C_{\bar P_2}-C_{\bar P_1}}{TV_{eff}(\beta_{2}-\beta_{1})}=\frac{\Delta C_{\bar P}}{T V_{eff} \Delta \beta},\label{5.1}\\
\left(\frac{\partial\bar P}{\partial T}\right)_{V_{eff}}&=&\frac{\beta_{2}-\beta_{1}}{\kappa_{T_2}-\kappa_{T_1}}=\frac{\Delta \beta}{\Delta \kappa_{T}}.\label{5.101}
\end{eqnarray}
where $\beta=\frac{1}{V_{eff}}\left(\frac{\partial V_{eff}}{\partial T}\right)_{\bar P}$ is the volume expansion coefficient and $\kappa_{T}=-\frac{1}{V_{eff}}\left(\frac{\partial V_{eff}}{\partial \bar P}\right)_{T}$ is the isothermal compressibility
coefficient.  For a genuine second order phase transition both of these equations have to be satisfied simultaneously. From eq.~(\ref{deltabarM}), we can obtain the following relation
\begin{equation}\label{5.2}
\left(\frac{\partial \bar P}{\partial T}\right)_{\bar S}=\left(\frac{\partial \bar S}{\partial V_{eff}}\right)_{\bar P}, \quad \left(\frac{\partial \bar P}{\partial T}\right)_{V_{eff}}=\left(\frac{\partial \bar S}{\partial V_{eff}}\right)_{T}.
\end{equation}
With the above equations, the Prigogine-Defay (PD) ratio $\Pi$ becomes
\begin{equation}\label{5.3}
\Pi=\left(\frac{\partial \bar P}{\partial T}\right)_{\bar S}/\left(\frac{\partial \bar P}{\partial T}\right)_{V_{eff}}=\left(\frac{\partial\bar S}{\partial V_{eff}}\right)_{\bar P}/\left(\frac{\partial \bar S}{\partial V_{eff}}\right)_{T}.
\end{equation}
The definition of the PD ratio was presented by Prigogine and Defay~\cite{Prigogine} and reviewed in Ref.~\cite{Prabhat}. At the critical point $(T_c, \bar P_{c}, V_{eff}^c)$, we have
\begin{equation}\label{5.4}
\left(\frac{\partial \bar P}{\partial V_{eff}}\right)_{T}=\left(\frac{\partial^{2} \bar P}{\partial V_{eff}^{2}}\right)_{T}=0.
\end{equation}
Substituting Eq.~(\ref{5.2}) into Eqs.~(\ref{5.1}) and~(\ref{5.101}), at the critical point we can obtain
\begin{equation}\label{5.5}
\frac{\Delta C_{\bar P}}{T_{c} V_{eff}^{c} \Delta \beta}=\left[\left(\frac{\partial \bar S}{\partial V_{eff}}\right)_{\bar P}\right]^{c}, \quad \frac{\Delta \beta}{\Delta \kappa_{T}}=\left[\left(\frac{\partial \bar S}{\partial V_{eff}}\right)_{T}\right]^{c}.
\end{equation}
On the other hand, since $\bar S=\bar S(\bar P,~V_{eff})$, therefore,
\begin{eqnarray}\label{5.6}
\left(\frac{\partial\bar S}{\partial V_{eff}}\right)_{T}=\left(\frac{\partial\bar S}{\partial \bar P}\right)_{V_{eff}}
\left(\frac{\partial\bar P}{\partial V_{eff}}\right)_{T}+\left(\frac{\partial\bar S}{\partial V_{eff}}\right)_{\bar P}.
\end{eqnarray}
From eq. (\ref{5.4}) we can find that, $\left(\frac{\partial\bar P}{\partial V_{eff}}\right)_{T}=0$ and $\left(\frac{\partial\bar S}{\partial \bar P}\right)_{V_{eff}}$ has a finite value at the critical point. Therefore the first term of the right side for above equation vanishes. That is very special thermodynamical feature of AdS black holes, which may not still hold on for other systems. Thus we have
\begin{eqnarray}\label{5.7}
\left[\left(\frac{\partial\bar S}{\partial V_{eff}}\right)_{T}\right]^{c}=\left[\left(\frac{\partial\bar S}{\partial V_{eff}}\right)_{\bar P}\right]^{c},
\end{eqnarray}
Then substituting Eq.~(\ref{5.7}) into Eq.~(\ref{5.3}), the universal PD ratio $(\prod)$ at the critical point becomes
\begin{equation}\label{5.8}
\prod =1.
\end{equation}
Hence from the PD ratio perspective, the phase transition occurring at $T=T_{c}$ is a second-order equilibrium transition as well as other AdS black holes \cite{Banerjee004,Banerjee156,Banerjee317}, which is also consistent with the results in the last section. In other words, the phase transition of AdS black holes is independent with the phase spaces, such as the extended phase space and the restricted phase space.

\section{Discussions and Conclusions}
\label{scheme7}
In this manuscript we have studied the thermodynamics of the EPYM AdS black hole in the restricted phase space which has revealed several remarkable characters as the same as the RN-AdS black hole, and compared them with that in the expanded phase space. The results will be summarized in the following
\begin{itemize}
\item{The first law of thermodynamics for the EPYM AdS black hole in the restricted phase space conforms to the standard description of ordinary thermodynamical systems: the mass parameter is to be understood as the internal energy, and the Euler relation of this system in the restricted phase space is restored as in an ordinary thermodynamical system;}
\item{In these two different phase spaces, the property of phase transition including the first-order and second-order phase transitions for the EPYM AdS black hole does not change. That means that the thermodynamical property of AdS black holes is independent with the adoption of corresponding phase spaces;}
\item{From the PD ratio perspective, this charged non-linear black hole is indeed in a equilibrium state at $T=T_c$ as well as ordinary thermodynamical systems. This is also indicating that black holes can be indeed regarded as thermodynamical systems.}
\end{itemize}

\section*{Acknowledgements}

We would like to thank Prof. Ren Zhao for his indispensable discussion and comment. This work was supported by the National Natural Science Foundation of China (Grant No. 12075143) and the Science Technology Plan Project of Datong City, China (Grant No. 2020153).

\end{document}